\documentstyle[preprint,aps]{revtex}
\draft
\begin{document}

\preprint{\vbox{\baselineskip = 14pt
\rightline{hep-ph 9503248}\break
\rightline{OCIP/C-94-11}\break
\rightline{November 1994}}}

\title{\bf Some constraints on neutral heavy leptons\\
from flavor-conserving decays of the $Z$ boson}
\author{G.\ Bhattacharya, Pat Kalyniak, and I.\ Melo}
\address{Ottawa-Carleton Institute for Physics\\
Department of Physics, Carleton University\\
1125 Colonel By Drive,
Ottawa, Ontario, Canada K1S 5B6}

\maketitle

\begin{abstract}
{Small neutrino masses can arise in some grand unified models or superstring
theories. We consider a model with an enhanced fermion sector containing Dirac
neutral heavy leptons. The dependence on the mass and mixing parameters of
these new fermions is investigated for several measurable quantities. We
study the flavor-conserving leptonic decays of the $Z$ boson and universality
breaking in these decays. We also consider the $W$ boson mass dependence on
neutral heavy lepton parameters.}
\end{abstract}

\medskip


\newpage
\section{Introduction}\label{intro}

Experimental evidence suggests that neutrino masses are very small, if not
zero. However, the way the standard model of electroweak interactions (SM)
accommodates massless neutrinos, by the absence of right-handed neutrino
fields, is considered unnatural. Interesting solutions to this problem have
been suggested in the low energy limit of superstring theories\cite{strings},
and in grand unified theories (GUT's)\cite{guts}.
Naturally small neutrino masses may even be accommodated within the framework
of
the $SU(2)_{L} \times U(1)_{Y}$ symmetry. Either the fermion content of the SM
or the Higgs sector may be extended. In
this work, we concentrate on the former option.

We consider an $SU(2)_{L} \times U(1)_{Y}$ based, superstring-inspired
model with an extended fermion sector including
neutral heavy leptons (NHL's)\cite{valle,bernabeu1}.
The NHL's are Dirac particles and B--L (baryon minus lepton number)
conservation is imposed as an unbroken symmetry. This is in
contrast to see-saw models \cite{guts,Ilakovac} wherein both light
neutrinos and NHL's are Majorana particles and B--L is broken.

The model considered allows for the possibilities of lepton-flavor
violation, universality violation, and CP violation. Our primary interest here
is in obtaining potential constraints on the model from current experimental
data on $Z$ leptonic decay widths. We focus on the direct contribution of NHL's
to flavor-conserving leptonic $Z$ decays via one-loop diagrams. In
addition to the $Z$ partial widths for individual lepton flavors, we also
calculate a measure of the leptonic universality violation, as defined in Ref.\
6. Further, we find that we must take into account the
impact of NHL's on the mass of the $W$ boson since $M_{W}$ is an input
parameter
within the renormalization scheme adopted here. Hence, we display also the
dependence of $M_{W}$ on the parameters of our model.

This paper is organized as follows. In Sec.\ \ref{model}, we describe the
model.
Existing experimental constraints on the parameters of the model, namely the
masses and mixings of the neutral leptons, are reviewed in Sec.\ \ref{rev}.
Our one-loop
calculation of the $Z$ leptonic decay, $Z \rightarrow l^{+}l^{-}$, is
presented in Sec.\ \ref{zlfc}. This section also contains a discussion of our
renormalization scheme, including the consideration of $M_{W}$. Many of the
detailed results are relegated to an Appendix. In Sec.\ \ref{results}, we
present our
results on the $Z$ leptonic widths, the universality violating measure, and the
$W$ mass. We summarize and draw our conclusions in the final section.

\section{Description of the model}\label{model}

Originally, only broken B--L symmetry and Majorana neutrinos were
thought of as providing an understanding of the smallness of neutrino masses.
However, as discussed in Ref.\ 4, superstring inspired models
can have small neutrino masses (in fact, zero) even if B--L symmetry is
unbroken and NHL's are Dirac particles. In these models,
the SM particle content is extended by two new neutrino fields,
$N_{R} (0,0)$ and $S_{L}(0,0)$, per family; the zeros indicate
$SU(2)_{L} \times U(1)_{Y}$ quantum numbers. Imposing total lepton number
conservation leads to the mass matrix
\begin{equation}
{\cal L}_{mass}  = -\frac{1}{2} (\bar{\nu_{L}} \bar{\hat{N_{L}}} \bar{S_{L}})
\left(
\begin{array} {lll}
               0     &  D  &  0        \\
               D^{T} &  0  &  M^{T}    \\
               0     &  M  &  0
\end{array}
\right)
\left(
\begin{array} {c}
\hat{\nu_{R}} \\
N_{R}        \\
\hat{S_{R}}
\end{array}
\right),
\label{eqn1}
\end {equation}
where $\nu_{L} = (\nu_{L}^{e}, \nu_{L}^{\mu}, \nu_{L}^{\tau})$ and
$\hat{\nu_{R}} = |CPT \rangle \nu_{L}$. D and M are $3 \times 3$ mass matrices.
The diagonalization of the mass matrix yields three massless neutrinos
($\nu_i$) along
with three Dirac NHL's ($N_a$) of mass $M_{N} \sim M$. Note that this
implies there are no time dependent neutrino oscillations and no neutrinoless
double beta decays. The weak eigenstates $\nu_{L}$ are mostly massless
neutrinos with a small mixing ($\sim D/M$) of NHL's. The NHL mixing in this
model is not restricted by small neutrino masses (as is often the case with
see-saw models where both the mixing and the masses of light neutrinos are
sensitive to the $D/M$ ratio), and hence rates for all interesting phenomena
can be large \cite{bernabeu1,Dittmar,ggjv}. This model is thus attractive not
only conceptually, but also practically.

The weak interaction eigenstates $\nu_{L}$ are related to six mass eigenstates
$n_{\alpha}$ via a $3 \times 6$ mixing matrix $K$ with components
$K_{ln_{\alpha}}$; $l = e, \mu, \tau$ and $n_{\alpha} = \nu_{1}, \nu_{2},
\nu_{3}, N_{4}, N_{5}, N_{6}$.
\begin{eqnarray}
  {\begin{array}{ll}
  K_{ln_{\alpha}}= & {\left( \begin{array}{llllll}
  K_{e\nu_{1}} & K_{e\nu_{2}} & K_{e\nu_{3}} &
  K_{eN_{4}}   & K_{eN_{5}}   & K_{eN_{6}} \\
  K_{\mu\nu_{1}} & K_{\mu\nu_{2}} & K_{\mu\nu_{3}} &
  K_{\mu N_{4}}   & K_{\mu N_{5}}   & K_{\mu N_{6}} \\
  K_{\tau\nu_{1}} & K_{\tau\nu_{2}} & K_{\tau\nu_{3}} &
  K_{\tau N_{4}}   & K_{\tau N_{5}}   & K_{\tau N_{6}} \\
\end{array}    \right)} \end{array}} & \equiv & (K_{L},\; K_{H}).
 \label{eqn2}
\end{eqnarray}
After rotating away redundant degrees of freedom from $K$, we are left with
$3^{2}$ angles and ${(3-1)}^{2}$ phases.
This allows for possible lepton-flavor violation, universality
violation and CP violation.

   The mixing factor which typically governs flavor-conserving
processes, is given by:

\begin{eqnarray}
ll_{mix} & = & \sum_{n_{\alpha}=N_{4},N_{5},N_{6}} {|K_{ln_{\alpha}}|}^{2},
\makebox[.5in] [c] { } l= e, \mu, \tau
\label{eqn3}
\end{eqnarray}
and the flavor-violating mixing factor $l_{a}{l_{b}}_{mix}$
is defined as:

\begin{eqnarray}
l_{a}{l_{b}}_{mix} & = & \sum_{n_{\alpha}=N_{4},N_{5},N_{6}} K_{l_{a}
n_{\alpha}} K_{l_{b} n_{\alpha}}^{\star},
\makebox[.5in] [c] { } l_{a,b} = e, \mu, \tau,
\makebox[.2in] [c] { }l_{a} \neq l_{b}.
\label{eqn4}
\end{eqnarray}

Further, an important inequality holds:

\begin{eqnarray}
|{l_{a}{l_{b}}_{mix}}^2| & \leq & {l_{a}l_{a}}_{mix}{l_{b}l_{b}}_{mix},
\makebox[.5in] [c] { } a \neq b.
\label{eqn5}
\end{eqnarray}
This implies  that one might observe nonstandard effects in flavor-conserving
processes even if they are absent in flavor-violating processes.

For reference, the charged current Lagrangian is given by

\begin{eqnarray}
{\cal L}_{cc}  =  \frac{1}{2 \sqrt{2}} g W^{\mu} \sum_{l=e, \mu ,\tau}
 \sum_{n_{\alpha}} \bar{l} \gamma_{\mu} (1-\gamma_{5}) K_{ln_{\alpha}}
 n_{\alpha};
\; \; n_{\alpha} = \nu_{1},\nu_{2},\nu_{3},N_{4},N_{5},N_{6}
\end{eqnarray}
and the neutral current Lagrangian as (the $ZNN$ part is obtained by analogy)
\begin{eqnarray}
{\cal L}_{nc}  = \frac{g}{4c_{W}} Z^{\mu} \sum_{i=1,2,3;a=4,5,6} \bar{\nu_{i}}
{(K_{L}^{\dagger}K_{H})}_{ia} \gamma_{\mu} (1-\gamma_{5})N_{a},
\end{eqnarray}
where $c_{W} = {\rm cos} \theta_{W}$, $\theta_{W}$ being the Weinberg angle.

\section{Review of Existing Constraints on Neutral Heavy Leptons}\label{rev}

Constraints on neutral heavy lepton masses and mixings come from three
different sources.
First, there is the possibility of direct production of NHL's. For instance,
if an NHL is light enough, it could be produced in some decays, {\sl e.g.}
$Z \rightarrow N_{a} + \nu$, and subsequently decay itself.
The rate for $Z$ decays into an NHL and a light neutrino has been given
previously\cite{Dittmar} as

\begin{eqnarray}
\Gamma(Z \rightarrow N_{a} + \nu) & = & a_{mix}
(1-\frac{{M_{N_{a}}}^{2}}{{M_{Z}}^{2}})(1+\frac{{M_{N_{a}}}^{2}}{{2
M_{Z}}^{2}})\Gamma(Z \rightarrow \nu + \nu)
\end{eqnarray}
where

\begin{eqnarray}
a_{mix} & = & \sum_{l=e,\mu,\tau} {|K_{lN_{a}}|}^{2}.
\end{eqnarray}
The subsequent NHL decay rate (for $M_{N} \leq M_{W}$) is then
given by

\begin{eqnarray}
\Gamma_{N} & = & a_{mix} (\frac{M_{N}}{m_{\mu}})^{5}
\Phi_{l}\Gamma_{\mu},
\end{eqnarray}
where $\Gamma_{\mu}$ is the muon decay rate and $\Phi_{l}$ is the
effective number of decay channels available to the NHL\cite{Gronau}.
Given the absence of experimental evidence for such direct production,
we will consider only NHLs with mass greater than
the $Z$ mass.

Secondly, there are constraints on NHL mixing parameters from a variety of low
energy experiments and from experiments at the CERN Large Electron Positron
Collider I (LEP I). Due to unitarity properties of the mixing
matrix $K$, a nonzero NHL mixing slightly reduces the couplings of light
neutrinos from their standard model values, thus affecting rates
for nuclear $\beta$ decays, $\tau$ and $\pi$ decays, and for $Z$ decays.
The following upper limits are consistent with experiment\cite{Nardi}

\begin{eqnarray}
      ee_{mix} & \leq & 0.0071 \nonumber \\
        \mu\mu_{mix} & \leq  & 0.0014 \\
         \tau\tau_{mix} & \leq & 0.033  \makebox[.9in][c]{or} \leq 0.024
 \makebox[1.5in][r]{including LEP I}\nonumber
 \end{eqnarray}
These limits are model independent and also independent of the NHL mass. They
arise from a global analysis of results including lepton universality
experiments, Cabibbo-Kobayashi-Maskawa (CKM) unitarity tests, $W$ mass
measurements and results from
LEP I  experiments. Note
that the LEP constraints presented above do not include NHL loop effects but,
rather, only coupling constant modifications due to mixing. We consider NHL
loop effects in this work.

Finally, the NHL masses and mixings can be constrained via their
contribution in loops to various processes. Such constraints are NHL mass
dependent. Flavor-violating processes have previously been studied at low
energies; these include
$\mu \rightarrow e \gamma$, $\mu \rightarrow 3 e$\cite{bernabeu1,ggjv,Ng}
and  flavor violating decays of the $\tau$\cite{ggjv,pilaftsis1}.
For instance, in the context of the model considered here, with mass degenerate
NHLs, the $\mu \rightarrow e \gamma$ branching ratio is
\cite{bernabeu1,ggjv}

\begin{eqnarray}
BR(\mu \rightarrow e \gamma) & = & \frac
{3\alpha}{32\pi}{|e\mu_{mix}|}^2 {|F_{\gamma}(x)|}^{2}
\end{eqnarray}
where $x = \frac{M_N^2}{M_W^2}$, and $F_{\gamma}(x)$ is an NHL mass dependent
form factor. For NHL masses
$M_{N} > 500$ GeV, which we will ultimately consider,
$F_{\gamma}(x) \rightarrow -2$. Given the current experimental limit on the
$\mu \rightarrow e \gamma$ branching ratio ($\leq$ 4.9 x $10^{-11}$)\cite{pdb},
this yields an upper limit on the mixing of

\begin{eqnarray}
|e\mu_{mix}| & \leq & 0.00024.
\end{eqnarray}
By combining the constraints obtained from the global analysis
(Eq.\ (11)) with the inequality relations of Eq.\ (\ref{eqn5}) one
obtains the following upper limits on the mixing factors

\begin{eqnarray}
|e\mu_{mix}| & \leq & 0.0032 \nonumber \\
|\mu\tau_{mix}| & \leq & 0.0068 \\
|e\tau_{mix}| & \leq & 0.015 \nonumber
\end{eqnarray}
For the mixings $\mu\tau_{mix}$ and $e\tau_{mix}$, these are the
strongest available constraints.
In addition, flavor-violating leptonic $Z$ decays,
$Z \rightarrow e\mu, e\tau, \mu\tau$, have also been studied
\cite{bernabeu1,Korner}.
In this work, we consider the flavor-conserving decays
$Z \rightarrow ee, \mu\mu, \tau\tau $.

\section{Calculation of the one-loop level contribution of Neutral Heavy
Leptons to Leptonic Flavor-Conserving $Z$ decays}\label{zlfc}

As noted previously, the limits on mixing parameters extracted
from LEP I observables\cite{Nardi}
do not include NHL's in the one-loop diagrams; only mixing factor
modifications
are made to the tree level results. We consider here the direct
contribution to flavour conserving leptonic $Z$ decay of NHL's via one-loop
diagrams. The importance of studying these processes is enhanced by the
implication of Eq.\ (\ref{eqn5}); one may observe flavor-conserving
(but universality breaking) $Z$ decays even in the absence of flavour
violating processes.

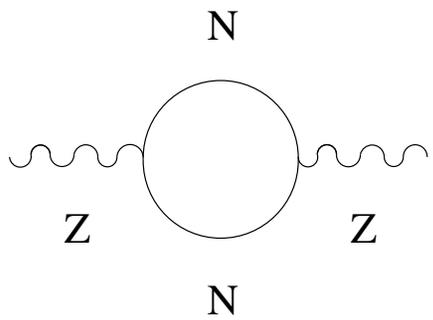
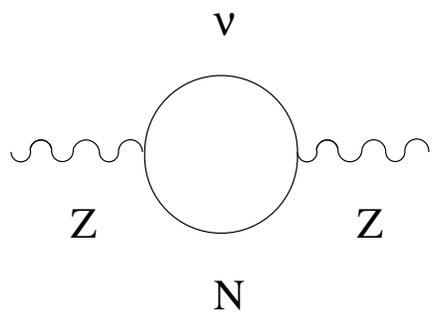
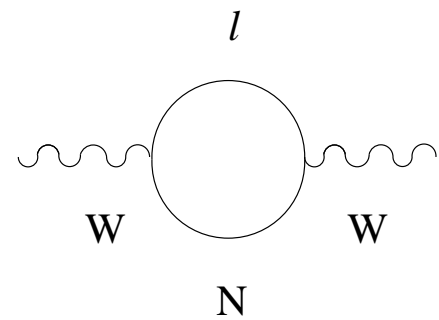
\begin{figure}
\caption{Diagrams for oblique corrections due to neutral heavy leptons N.}
\label{fig1}
\end{figure}

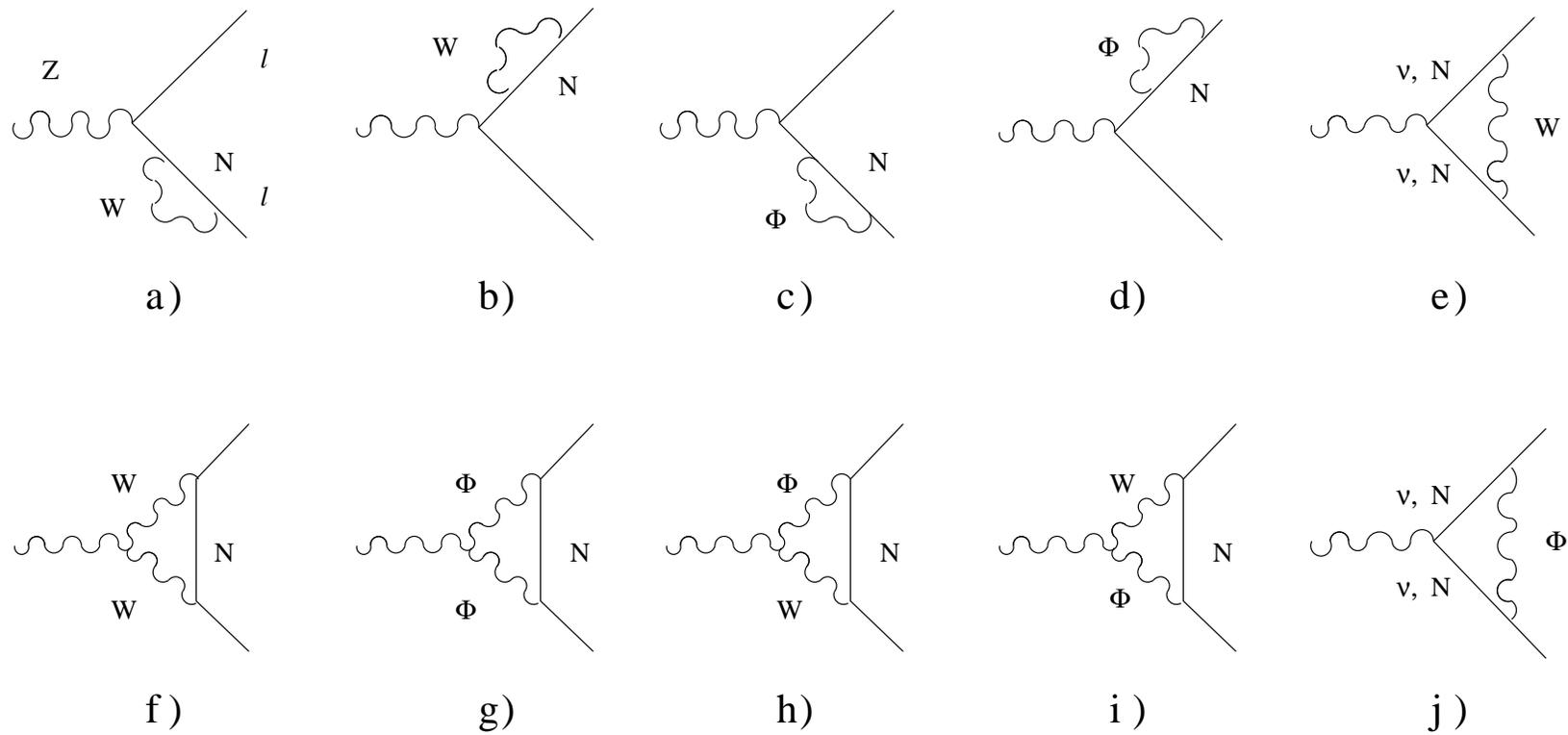
\begin{figure}
\caption{Diagrams for one-loop vertex correction to flavor-conserving
leptonic $Z$ decays due to neutral
heavy leptons N.}\label{fig2}
\end{figure}

For $Z$ leptonic decay, NHLs contribute directly to $Z$ oblique corrections, as
shown in Fig.\ \ref{fig1}, to lepton wave function renormalizations and to
vertex corrections, as in Fig.\ \ref{fig2}. These one-loop contributions of
NHL's can be
incorporated into the framework of the full standard model one-loop electroweak
corrections. The one-loop corrected leptonic width can be parametrized as
\cite{Hollik}

\begin{eqnarray}
\Gamma_{Z} &= & \frac{\Gamma_{0} + \delta\Gamma_{Z}}{1+\hat\Pi_{Z}(M_{Z}^{2})}
(1+\delta_{QED}),
\end{eqnarray}
where the tree level leptonic width of the $Z$ boson is given by

\begin{eqnarray}
\Gamma_{0} & = & \frac{\alpha}{3} M_{Z}(v_{f}^{2} + a_{f}^{2}),
\end{eqnarray}
$v_f$ and $a_f$ being respectively the vector and axial vector couplings
of charged leptons to $Z$.
The one-loop electroweak corrections include $\delta\Gamma_{Z}$, which
represent vertex loops, and $\hat\Pi_{Z}$ and $\delta_{QED}$ which
represent the $Z$-oblique corrections and QED corrections respectively.

Our calculation is done within the framework of an on-shell renormalization
scheme as detailed in Ref. 15. All the SM one-loop diagrams were calculated
using standard routines from the CERN electroweak library\cite{Hollik}
modified by appropriate mixing factors. The
vertex parameter $\delta\Gamma_Z$ actually includes also fermion wave function
renormalization and counterterm contributions in the scheme we adopt. This
on-shell renormalization scheme takes $\alpha, M_{Z} $ and $M_{W}$ as input
parameters. However, the direct measurement of $M_{W}$ is not
yet precise
enough for its use as an input parameter; hence, it is replaced by $G_{\mu}$
via the one-loop relation

\begin{eqnarray}
M_{W}^{2} s_{W}^{2} & = & \frac{\pi\alpha}{\surd 2 G_{\mu} (1-\Delta r)}
(1-\frac{1}{2}ee_{mix}-\frac{1}{2}\mu\mu_{mix}),
\end{eqnarray}
where $\Delta r$ is an equivalent of the SM quantity $\Delta r^{SM}$, and
$s_{W} = {\rm sin}\theta_{W}$.
As a result, we also have to consider muon decay loops with NHL's. The types of
corrections involved in muon decay are pictured schematically in Fig. 3. They
include boxes and vertices with NHL's as well as the lepton wave function
renormalizations and $W$ oblique corrections. The NHL diagrams contributing to
the $W$
oblique corrections are shown in Fig. 1 along with the $Z$ oblique corrections.
These corrections all feed into the $Z$ leptonic decay calculation indirectly
via
the dependence of $M_{W}$ on the parameter $\Delta r$ and the overall factor
modified by mixings.

Referring to Eq. (15) now, the QED corrections, $\delta_{QED}$, are not
modified from the SM. The factor ${(1 + \hat\Pi_{Z}(M_{Z}^{2}))}^{-1} $
represents the wave-function
renormalization of the $Z$ boson. It depends on all the unrenormalized
propagator corrections $(\Sigma_{Z}$ , $\Sigma_{W}$ , $\Sigma_{\gamma Z},
\Sigma_{\gamma})$ \cite{Hollik}. Of
these, $\Sigma_{\gamma Z}$ and $\Sigma_{\gamma}$ are not modified
within our model while
$\Sigma_{Z}$, $\Sigma_{W}$ both contain nonstandard terms.
Those nonstandard terms, denoted as $\Sigma_{Z}^{\nu,N}(s)$ and
$\Sigma_{W}^{\nu,N}(s)$,
are given by Eqs. (A2) and (A4) respectively, in the Appendix. They consist
of  $M_{N}$ dependent
terms representing the direct contribution from NHL's in loops in Fig. 1 and
of SM terms modified by mixing factors (the indirect effect of NHL's reducing
the mixings of light neutrinos through the unitary matrix $K$).

The vertex parameter $\delta \Gamma_{Z}$ includes $\gamma - Z$ mixing,
external fermion wave function renormalization, and counterterm contributions
in addition to the vertex loops involving NHL's. Those fermion self energy and
$Zf \bar{f}$ vertex loops which contain NHLs are shown in Figs. 2a-j.
The individual
contributions of Figs. 2a-j are given in Eq. (A6) in the Appendix.

It is characteristic that NHL's in loops generally do not decouple (violation
of
the Appelquist-Carazzone theorem)\cite{ac}; rather they often
show a quadratic mass dependence $\sim \frac{M_{N}^{2}}{M_{W}^{2}}$. This is a
common feature for theories based on the spontaneous symmetry breaking
mechanism.
We are already familiar with a similar result for the top quark mass in SM
loops\cite{bernabeu3}.
Indirect bounds on the top quark mass arise from the $Z$ and $W$ polarization
diagrams
and $Zb \bar{b}$ vertex loop since these corrections come in with the amplitude
\begin{eqnarray}
{\cal M}_{t} & \sim & {|V_{tb}|}^{2}\frac{m_{t}^{2}}{M_{W}^{2}}  \doteq
\frac{m_{t}^{2}}{M_{W}^{2}}.
\end{eqnarray}
While the top quark mixes with the full strength (${|V_{tb}|}^{2} \sim 1$),
the NHL mixings are limited by Eqs. (11) and (14).
As a result, we find sensitivity only
for $M_{N} \sim 10\: m_{t}$. Thus we will
only present numerical results for NHL masses greater than about
500 GeV.

Given the nondecoupling feature, the vertex correction
$\delta \Gamma_Z$ is dominated for large NHL masses by the diagrams
(in decreasing order of importance) 2j, 2f and 2e, while 2a-b, 2c-d, 2g, 2h
and 2i
are negligible (the largest unrenormalized contribution comes from 2g and 2c-d;
however, diagrams 2c-d enter the renormalized vertex correction
$\delta \Gamma_Z$ as a part
of a counterterm that cancels out the large amplitude of the graph 2g).

To illustrate how $\hat\Pi_Z$ and $\Delta r$ depend on $M_{N}$ (for $M_{N}$
large),
we separate $M_{N}$ dependent terms as $\hat\Pi_{M_{N}}$, $\Delta r_{M_{N}}$
and find in the limit ${\cal X}^{-1} = \frac{M_{W}^{2}}{M_{N}^{2}} \rightarrow
 0$ (large NHL mass)
\begin{eqnarray}
\Delta r_{M_{N}} & = & - \frac{\alpha}{\pi s_{W}^{4}}\biggl[
l_{HH}\frac{c_{W}^{2}}{16}{\cal X} +
(l_{LL}-1)\frac{1}{24}\sum_{l=e,\mu,\tau}\ln\frac{M_{N}^{2}}{m_{l}^{2}}\biggr]
\nonumber \\
\hat\Pi_{M_{N}} & = & \frac{\alpha}{\pi}\biggl[
l_{HH} \frac{c_{W}^{2}-s_{W}^{2}}{16s_{W}^{4}}{\cal X}
+ (l_{LL}-1)\frac{1}{24s_{W}^{4}}\sum_{l=e,\mu,\tau}
\ln\frac{M_{N}^{2}}{m_{l}^{2}}\biggr],
\end{eqnarray}
where

\begin{eqnarray}
l_{HH} & = & \sum_{l=e,\mu,\tau}\left(|le_{mix}|^2 + |l\mu_{mix}|^2 +
             |l\tau_{mix}|^2\right) \nonumber \\
l_{LL} & = & 1 - 2\left(ee_{mix}+\mu\mu_{mix}+\tau\tau_{mix}\right) + l_{HH}
\end{eqnarray}

With the NHL mass of the order of several TeV, one has to worry
about the
perturbative unitarity bound. A good way to demonstrate this is to bring about
the Higgs analogy. The width of the Higgs boson of mass $m_H$ is given by

\begin{eqnarray}
\Gamma_{H} & = & \frac{3 \alpha}{32 M_{W}^{2} s_{W}^{2}} m_{H}^{3},
\end{eqnarray}
which can be compared with the width of the NHL.
For NHL mass $M_{N} \gg M_{W}, M_{Z}, M_{H}$, the width
is\cite{pilaftsis2}:

\begin{eqnarray}
\Gamma_{N} & = & \frac{\alpha}{4M_{W}^{2}s_{W}^{2}}M_{N}^{3}a_{mix}.
\end{eqnarray}
Demanding that $\Gamma_{H} \leq \frac{1}{2} M_{H}$
we get the well known bound on the Higgs mass $M_{H} \leq
1 \; {\rm TeV}$
Similarly, demanding $\Gamma_{N} \leq \frac{1}{2} M_{N}$ for the NHL,
with the current constraints on mixings, (Eq.\ (11)), one obtains
$M_{N} \leq 3.5$ TeV.

\section{Results}\label{results}

In this Section, we present our numerical results. As input parameters, we used
$M_{Z} = 91.173$ GeV, $M_{H} =  200$ GeV,
$\alpha ^{-1} = 137.036$ and
$A \equiv \frac{\pi \alpha}{\sqrt{2} G_{\mu}} = 37.281 \: {\rm GeV}$.
We have assumed degenerate masses for the three NHL's and present results for
the NHL mass range $0.5 \: {\rm TeV} \leq M_{N} \leq 5$
TeV, as motivated
by the non-decoupling and perturbative unitarity arguments given in the last
Section. We have also imposed restrictions on the mixing parameters. We assume
that $ee_{mix}$ and $\mu \mu_{mix}$ are very small relative to $\tau
\tau_{mix}$. The model and NHL mass independent limits quoted in Eq. (11) are
more stringent for $e$ and $\mu$ than for $\tau$. In addition, our assumption
is also
partially supported by the smallness of $e\mu_{mix}$, as determined from
$\mu \rightarrow e \gamma$, in combination with the inequality Eq. (5). This
neglect of $ee_{mix}$ and $\mu \mu_{mix}$ proves useful practically in that
many of the muon decay loops (boxes and vertex corrections, but not $W$ oblique
correction) are eliminated as a result.

The $Z$ leptonic width is given as a function of NHL mass in Figs. 4a, b.
In Fig. 4a, we have fixed the mixing $\tau\tau_{mix} = 0.033$. The width for
$Z$ decay to $e^{+}e^{-}$ is shown for a top quark mass of 174 GeV. The
$Z$ decay rate into $\tau^{+}\tau^{-}$ is shown for three values of the top
quark mass, 150, 174 and 200 GeV. The dashed lines represent the $1\sigma$
variation about the current experimental result for the average $Z$-leptonic
width of $\Gamma_{l} = 83.96 \pm 0.18$ MeV\cite{zwidth}. In Fig. 4b, we fix the
top quark mass at 174 GeV and show the $Z$ width to $\tau^{+}\tau^{-}$
for three values of the mixing parameter, $\tau\tau_{mix} = 0.02, 0.033, 0.07$.
We also present results for the universality breaking ratio defined
as\cite{bernabeu2}
\begin{eqnarray}
U_{br} & = &
\displaystyle \left| \frac{\Gamma (Z \rightarrow \tau^{+} \tau^{-}) -
\Gamma (Z \rightarrow e^{+} e^{-})}
{\Gamma (Z \rightarrow \tau^{+} \tau^{-}) + \Gamma (Z \rightarrow e^{+}e^{-})}
              \right|.
\end{eqnarray}
This is shown in Fig. 5 as a function of $M_{N}$, again with $m_{t} = 174
\:$ GeV, and the mixing parameter varied about $\tau\tau_{mix} =
0.033$.
The $1\sigma$ experimental limit on $U_{br}$ is indicated as the dashed line.
Note that the most recently reported values of $Z$ widths into individual
lepton flavors\cite{zwidth} have $\Gamma (Z \rightarrow \tau^{+} \tau^{-}) >
\Gamma (Z \rightarrow e^{+} e^{-}), \; \Gamma (Z \rightarrow \mu^{+} \mu^{-})$,
as opposed to the last round of results\cite{pdb}.

Finally, we present the NHL mass dependence of the $W$ mass in Figs. 6a, b.
The top quark mass is varied in Fig. 6a, while the mixing is held constant at
$\tau\tau_{mix} = 0.033$. In Fig. 6b, the mixing is varied about
$\tau\tau_{mix} = 0.033$ for a fixed top quark mass $m_{t} = 174$
GeV.

\section{Discussion and Conclusions}\label{conc}

Our primary consideration here has been the inclusion of neutral heavy
leptons in the calculation of the flavor-conserving $Z$ decays to
charged leptons at one-loop level. The dependence of the $Z$ leptonic
widths on the NHL mass, $M_N$, and on the mixing parameter $\tau \tau_{mix}$
which we retain, was given in Figs. 4 a,b. We see for the experimentally
allowed upper limit of $\tau \tau_{mix} =$ 0.033, and assuming a top quark mass
$m_{t} =$ 174 GeV, the $Z$ decay width to $\tau$ leptons is
sensitive at the
present 2$\sigma$ level to NHL masses larger than about 2.5 TeV.
The top mass dependence is also shown in that Figure. The sensitivity to
$M_N$ and $m_t$ arises since these heavy fermions generally do not decouple
from the one-loop diagrams. Fig. 4 b indicates how the $Z$ width dependence
on $M_N$ varies with the mixing parameter. Apart from this comparison of each
leptonic width prediction with experiment we can also exploit the flavor
universality violation which takes place in the model. The universality
breaking ratio, $U_{br}$, defined in Sec. 4, is sensitive to NHL masses
above approximately 3.5 TeV at the 1$\sigma$ level, assuming
$\tau \tau_{mix} =$ 0.033.

The $W$ boson mass also exhibits some sensitivity to NHL parameters
arising from the mixing factor modifications and the presence of one-loop
diagrams containing neutral heavy leptons, as described in Sec. 3. From
Figs. 5a, b we see that the $W$ mass, currently measured as
$M_{W} = 80.23 \pm 0.18$ GeV\cite{mw1}, is sensitive at the
$1\sigma$ level
to NHL masses greater than about 3.5 TeV, again assuming
$\tau \tau_{mix} =$
0.033 and $m_{t} =$ 174 GeV. The experimental error on $M_W$ might
be expected
to come down to about 0.05 GeV once LEP II measures $W$ pair
production\cite{mw2}.

We have considered a model containing isosinglet neutral heavy leptons
which can accommodate various phenomena beyond the standard model, such as
lepton flavor-violation, CP violation and lepton universality violation.
We have presented the dependence on the mass and mixing parameters of this
model for $Z$ decays to
charged leptons and for the $W$ boson mass.
Because the NHL mass and mixing dependence is different for the $Z$ decay width
and the $W$ boson mass, they provide somewhat
complementary information on these parameters.

Current data from LEP I on $Z$ leptonic widths and the present Collider
Detector at Fermilab (CDF) and DO Collaboration
measurements of $M_W$ are sensitive to NHL masses greater than about
2.5 -- 3.5 TeV. With the accumulation of about 60
$({\rm pb})^{-1}$ at LEP I
in 1994 and the prospect of the very precise $W$ mass measurement at
LEP II, these sensitivities will certainly be improved considerably. Thus the
$Z$ partial width to leptons and $W$ mass measurements can provide, along
with the other observables discussed in Sec. 3, a consistency check on the
possible existence of isosinglet neutral heavy leptons.

\acknowledgements

This work was funded in part by the Natural Sciences and Engineering
Research Council of Canada. We would like to acknowledge useful
communications with M. C. Gonzalez-Garcia and A. Pilaftsis.

\appendix
\section*{}

In this Appendix, we present the parts of our calculation which are necessary
to obtain the NHL dependent modifications of the one-loop $Z$ decay width to
charged leptons. We refer the reader to Ref. 15 for further details.
We define the following symbols and mixing factors:
\begin{eqnarray}
{\cal X} & = & \frac{M_{N}^{2}}{M_{W}^{2}}, \nonumber \\
\Delta  & = & \frac{2}{\epsilon}-\gamma+\ln 4\pi + \frac{3}{2}\ln \mu^{2},
 \nonumber \\
\Delta^{m} & = & \frac{2}{\epsilon }-\gamma +\ln 4\pi -
\ln\frac{m^{2}}{\mu^{2}}, \nonumber \\
l_{1} & = & ll_{mix}, \nonumber \\
l_{2} & = & |le_{mix}|^2 + |l\mu_{mix}|^2
 + |l\tau_{mix}|^2, \nonumber \\
l_{3} & = & ll_{mix}-l_{2}, \nonumber \\
l_{4} & = & 1 - 2 ll_{mix} + l_{2}, \nonumber \\
l_{CH} & = & ee_{mix} + \mu \mu_{mix} + \tau \tau_{mix}, \nonumber \\
l_{HH} & = & \sum_{l=e,\mu,\tau}l_{2}, \nonumber \\
l_{LH} & = & 2 (l_{CH} - l_{HH}), \nonumber \\
l_{LL} & = & 1 - 2 l_{CH} + l_{HH}. \\     \nonumber
\end{eqnarray}
First we deal with the direct contributions of NHL's to the $Z$ boson wave
function renormalization, as parametrized by $\hat \Pi_{Z}(M_{Z}^{2})$.
$\hat \Pi_{Z}(M_{Z}^{2})$ depends on two quantities which are modified from
the SM
by the inclusion of NHL's. The neutral lepton part of the unrenormalized
self energy of the $Z$ boson, which consists of the SM massless $\nu$ loop, the
mixed $\nu N$ loop and the $NN$ loop, is given by
\begin{eqnarray}
\Sigma^{\nu ,N}_{Z}(s) & = &\frac{\alpha}{8\pi s_{W}^{2} c_{W}^{2}}\Biggl\{
\frac{s}{3}l_{LL}
\sum_{l=e,\mu,\tau}\biggl[\Delta^{m_{l}}+2-\ln(-\frac{s}{m_{l}^{2}}-i\epsilon)
-\frac{1}{3}\biggr] \Biggr. \nonumber \\
& & + \; l_{LH}\biggl[\Delta^{M_{N}}(\frac{s}{3}-\frac{M_{N}^{2}}{2})
+\frac{2}{9}s-\frac{M_{N}^{2}}
{6}+F(s,0,M_{N})(\frac{s}{3}-\frac{M_{N}^{2}}{6}-\frac{M_{N}^{4}}{6s})\biggr]
\nonumber \\
& & + \; l_{HH}\Biggl.
\biggl[\Delta^{M_{N}}(\frac{s}{3}-M_{N}^{2})-\frac{s}{9}+
\frac{1}{3}F(s,M_{N},M_{N})
(s-M_{N}^{2})\biggr]\Biggr\} ,
\end{eqnarray}
where the function $F(s,m_{1},m_{2})$ is given by
\begin{eqnarray}
F(s,m_{1},m_{2}) & = & -1 + \frac{m_{1}^{2}+m_{2}^{2}}{m_{1}^{2}-m_{2}^{2}}
\ln\frac{m_{1}}{m_{2}} \nonumber \\
& & - \int_{0}^{1} dx \ln \frac{sx^{2} -
x(s+m_{1}^{2}-m_{2}^{2})+m_{1}^{2}-i\epsilon}{m_{1}m_{2}}. \\
\nonumber
\end{eqnarray}
The leptonic part of the unrenormalized self energy of the $W$ boson is given
by:
\begin{eqnarray}
\Sigma^{\nu ,N}_{W}(s) & = &\frac{\alpha}{12\pi s_{W}^{2}}\Biggl\{
\sum_{l=e,\mu,\tau} (1 - l_{1})\biggl[
(s-\frac{3}{2}m_{l}^{2})\Delta^{m_{l}} \Biggr. \biggr.
\nonumber \\
& & + \biggl.
 (s-\frac{m_{l}^{2}}{2}-
\frac{m_{l}^{4}}{2s})F(s,0,m_{l})+\frac{2}{3}s-\frac{m_{l}^{2}}{2}\biggr]
 \nonumber
\\
& & + \sum_{l=e,\mu,\tau} l_{1} \biggl[ \frac{\Delta^{M_{N}}}{2}(s-\frac{5}{2}
M_{N}^{2}-
\frac{m_{l}^{2}}{2})+\frac{\Delta^{m_{l}}}{2}(s-\frac{5}{2}m_{l}^{2}-
\frac{M_{N}^{2}}{2}) \biggr. \nonumber \\
& & + \; (s-\frac{M_{N}^{2}+m_{l}^{2}}{2}-\frac{(M_{N}^{2}-m_{l}^{2})^{2}}{2s})
F(s,M_{N},m_{l}) \nonumber \\
& & + \Biggl.\biggl.(s-\frac{M_{N}^{2}+m_{l}^{2}}{2})(1-\frac{M_{N}^{2}+m_{l}^
{2}}{M_{N}^{2}-m_{l}^{2}}\ln\frac{M_{N}}{m_{l}})-\frac{s}{3}
\biggr] \Biggr\} \\  \nonumber
\end{eqnarray}

Next, we consider the contributions of NHL's to the parameter $\delta
\Gamma_{Z}$. The direct contribution of NHL's in the triangle diagrams of Figs.
2e-j is given by the sum of amplitudes:
\begin{eqnarray}
{\cal M} & = & +ie\epsilon_{\mu}\gamma^{\mu}(1-\gamma_{5})\frac{\alpha}{4\pi}
\Bigl\{ l_{1}{\cal M}_{Z\Phi W} + l_{2}{\cal M}_{ZNN \Phi} +
l_{1}{\cal M}_{Z\Phi \Phi} + (1-l_{1}){\cal M}_{ZWW\nu} \Bigr. \nonumber \\
& &
+ \Bigl. l_{1}{\cal M}_{ZWWN} + l_{3}{\cal M}_{ZN\nu W} +
l_{3}{\cal M}_{Z\nu NW}
+ l_{4}{\cal M}_{Z\nu \nu W} + l_{2}{\cal M}_{ZNNW}\Bigr\}, \\ \nonumber
\end{eqnarray}
with
($\frac{m_{l}^{2}}{M_{W}^{2}}$ terms neglected) :
\begin{eqnarray*}
{\cal M}_{Z\Phi W} & = & +\frac{ M_{W}^{2}}
{2 s_{W}c_{W}}{\cal X}C_{0}(M_{W},M_{N},M_{W}), \nonumber \\
{\cal M}_{ZNN \Phi} & = & +
\frac{ M_{W}^{2}}{8 s_{W}^{3} c_{W}}{\cal X}^{2}
C_{0}(M_{N},M_{W},M_{N}), \nonumber \\
{\cal M}_{Z\Phi \Phi} & = & - \frac{1}
{2 s_{W}^{3}} \frac{1-2s_{W}^{2}}{2c_{W}}{\cal
X}[C_{24}^{fin}(M_{W},M_{N},M_{W}) + \frac{1}{4}\Delta], \nonumber \\
\end{eqnarray*}
\newpage
\begin{eqnarray}
{\cal M}_{ZWWa} & = & -  \frac{3 c_{W}}
{4 s_{W}^{3}}\left\{\frac{2}{3}M_{Z}^{2}[-C_{11}(M_{W},M_{a},M_{W})
\right. \nonumber \\
& & -C_{23}(M_{W},M_{a},M_{W}) -C_{0}(M_{W},M_{a},M_{W})] \nonumber \\
& & \left. +4C_{24}^{fin}(M_{W},M_{a},M_{W})-\frac{2}{3}+ \Delta\right\},
\nonumber \\
{\cal M}_{ZabW} & = & -\frac{1}
{8 s_{W}^{3} c_{W}}\left\{2M_{Z}^{2}[(C_{23}(M_{a},M_{W},M_{b}) \right.
\nonumber \\
&   & \left. + C_{11}(M_{a},M_{W},M_{b})]+2-4C_{24}^{fin}(M_{a},M_{W},M_{b})
-\Delta \right\}
\nonumber \\
&   & \mbox{where a,b run over }  N, \nu \mbox{ ; } M_{\nu} = 0.
 \\  \nonumber
\end{eqnarray}
Here ${\cal M}_{Z\Phi W}$ is the sum of equal contributions from diagrams 2h
and 2i.
Diagram 2f comes in both with massless $\nu$'s and NHL's. Diagram 2e comes in
with four combinations of neutral lepton types. Thus ${\cal M}_{ZWW\nu }$ and
${\cal M}_{Z\nu \nu W}$ are standard model results but come into the sum (Eq.
(A5)) with NHL mixing factor coefficients. Our results in Eq. (A5) are written
in terms of the 't Hooft-Veltman integrals\cite{thooft}. Our conventions are
given below, with finite parts indicated by the superscript.

The function $C_{0}$ is defined as:
\begin{eqnarray}
C_{0}(m_{1},m_{2},m_{3}) & = & C_{0}(p_{1},p_{2};m_{1},m_{2},m_{3})
= C_{0}^{fin}(m_{1},m_{2},m_{3}) \nonumber
\\
& = &  -\int \frac{d^{n}q}{i\pi^{2}}
\frac{1}{D}, \nonumber \\
\mbox{where  } D & = & (q^{2}-m_{1}^{2}+i\epsilon)[(q-p_{1})^{2}-m_{2}^{2}
+i\epsilon] \nonumber \\
& & \times
[(q-p_{1}-p_{2})^{2}-m_{3}^{2}+i\epsilon].  \\   \nonumber
\end{eqnarray}
The functions  $C_{24}, C_{23}, C_{11}$ are defined by:
\begin{eqnarray}
C_{\mu} & = & -\int \frac{d^{n}q}{i\pi^{2}}\frac{q_{\mu}}{D} = -p_{1\mu}C_{11}
-
p_{2\mu}C_{12}, \nonumber \\
C_{\mu\nu} & = & -\int \frac{d^{n}q}{i\pi^{2}} \frac{q_{\mu}q_{\nu}}{D}
\nonumber \\
& = & p_{1\mu}p_{1\nu}C_{21} + p_{2\mu}p_{2\nu}C_{22} + (p_{1\mu}p_{2\nu} +
p_{1\nu}p_{2\mu})C_{23} - g_{\mu\nu}C_{24}.
\end{eqnarray}
The functions  $C_{24}, C_{23}, C_{11}$ are reduced (in the limit
$p_{1}^{2} = p_{2}^{2} = m_{l}^{2} \ll  (p_{1}+p_{2})^{2}=M_{Z}^{2}$) to:
\begin{eqnarray}
C_{24}(m_{1},m_{2},m_{3}) & = & \frac{1}{4}(\frac{2}{4-n}-\gamma-\ln \pi) +
 C_{24}^{fin}(m_{1},m_{2},m_{3}), \nonumber \\
C_{24}^{fin}(m_{1},m_{2},m_{3}) & = & [m_{1}^{2}C_{0}(m_{1},m_{2},m_{3}) +
f_{1}C_{11}(m_{1},m_{2},m_{3}) + \nonumber \\
& &B_{1}^{fin}(p_{1}+p_{2};m_{1},m_{3})](-\frac{1}{2}) + \frac{1}{4}
 \nonumber \\
\mbox{with  } f_{1} & = & m_{1}^{2}-m_{2}^{2}, \nonumber \\
C_{11}(m_{1},m_{2},m_{3}) & = & C_{11}^{fin}(m_{1},m_{2},m_{3}) =
 -\frac{1}{M_{Z}^{2}}[f_{2}C_{0}(m_{1},m_{2},m_{3}) - \nonumber \\
& &
 B_{0}^{fin}(p_{1}+p_{2};m_{1},m_{3}) + B_{0}^{fin}(p_{1};m_{1},m_{2})]
 \nonumber \\
\mbox{with  } f_{2} & = & M_{Z}^{2}+m_{2}^{2}-m_{3}^{2}, \nonumber \\
C_{23}(m_{1},m_{2},m_{3}) & = &  C_{23}^{fin}(m_{1},m_{2},m_{3}) =
 -\frac{1}{M_{Z}^{2}}
[B_{1}^{fin}(p_{1}+p_{2};m_{1},m_{3}) + B_{0}^{fin}(p_{2};m_{2},m_{3})
\nonumber \\
& & + f_{1}C_{11}(m_{1},m_{2},m_{3})] + C_{24}^{fin}(m_{1},m_{2},m_{3})
 \frac{2}{M_{Z}^{2}}. \\                                  \nonumber
\end{eqnarray}
The functions $B_{0}, B_{1}$ are defined as:
\begin{eqnarray}
B_{0}(p;m_{1},m_{2}) & = & \int \frac{d^{n}q}{i\pi^{2}}
\frac{1}{(q^{2}-m_{1}^{2}+i\epsilon)[(q-p)^{2}-m_{2}^{2}+i\epsilon]} \nonumber
\\
& = & (\frac{2}{4-n} -\gamma -\ln \pi) + B_{0}^{fin}(p;m_{1},m_{2}),
 \nonumber \\
B_{0}^{fin}(p;m_{1},m_{2}) & = & -\int_{0}^{1} dx \ln [p^{2}x^{2} + m_{1}^{2}
-(p^{2}+m_{1}^{2}-m_{2}^{2})x], \nonumber \\
B_{1}(p;m_{1},m_{2}) & = & -\frac{1}{2}(\frac{2}{4-n}-\gamma-\ln \pi)
+ B_{1}^{fin}(p;m_{1},m_{2}), \nonumber \\
B_{1}^{fin}(p;m_{1},m_{2}) & = & \int_{0}^{1} dx \ln [p^{2}x^{2} + m_{1}^{2}
-(p^{2}+m_{1}^{2}-m_{2}^{2})x] x.         \\ \nonumber
\end{eqnarray}

The remaining diagrams of Fig. 2 contribute NHL dependent terms to the charged
lepton wave function renormalization. The contributions to the lepton self
energy of Figs. 2a, b and Figs. 2c, d are given as $\Sigma ^{W}$ and $\Sigma
^{\Phi}$, respectively.
\begin{eqnarray}
-i\Sigma^{W} & = & \frac{i\alpha}{32\pi s_{W}^{2}}l_{1}
[2-2\ln 4\pi -2\ln \mu^{2} + 2\gamma -\frac{4}{\epsilon} +
4f({\cal X})]p_{\alpha}\gamma^{\alpha}(1-\gamma_{5}), \nonumber \\
-i\Sigma^{\Phi} & = & - \frac{i\alpha}{32\pi s_{W}^{2}}l_{1} {\cal X}
[ \frac{2}{\epsilon} + \ln 4\pi +\ln \mu^{2} -\gamma - 2f({\cal X})]
p_{\alpha}\gamma^{\alpha}(1-\gamma_{5}), \nonumber \\
\mbox{where } f({\cal X}) & = & \frac{{\cal X}^{2}}{2({\cal X}-1)^{2}}\ln {\cal
X} + \frac{{\cal X}}{1-{\cal X}}
-\frac{{\cal X}+1}{4(1-{\cal X})} + \frac{1}{2} \ln M_{W}^{2}.  \\  \nonumber
\end{eqnarray}
All these contributions, Eq. (A5) and (A11), along with their corresponding
counterterms modify the $Zl^{+}l^{-}$ vertex. In addition, the vertex is
modified via $\gamma-Z$ mixing. However, the relevant term $\Pi^{\gamma
Z}(M_{Z}^{2})$ depends on NHL's only through the $Z$ and $W$ self energies so
we
need no other results to determine $\delta \Gamma_{Z}$.

The results presented in this Appendix are also sufficient to derive the $W$
oblique corrections of $\Delta r$, which we have considered.

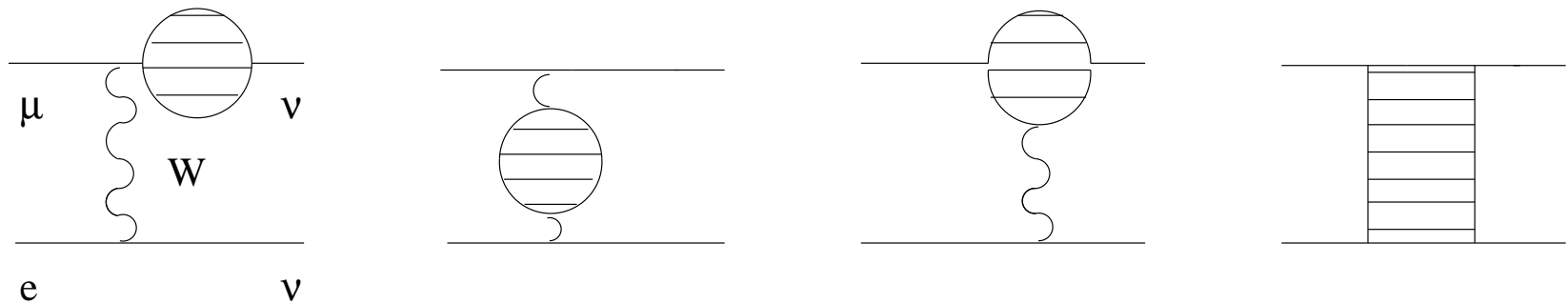
\begin{figure}
\caption{Schematic representation of the one-loop muon decay diagrams with
neutral heavy leptons}\label{fig3}
\end{figure}

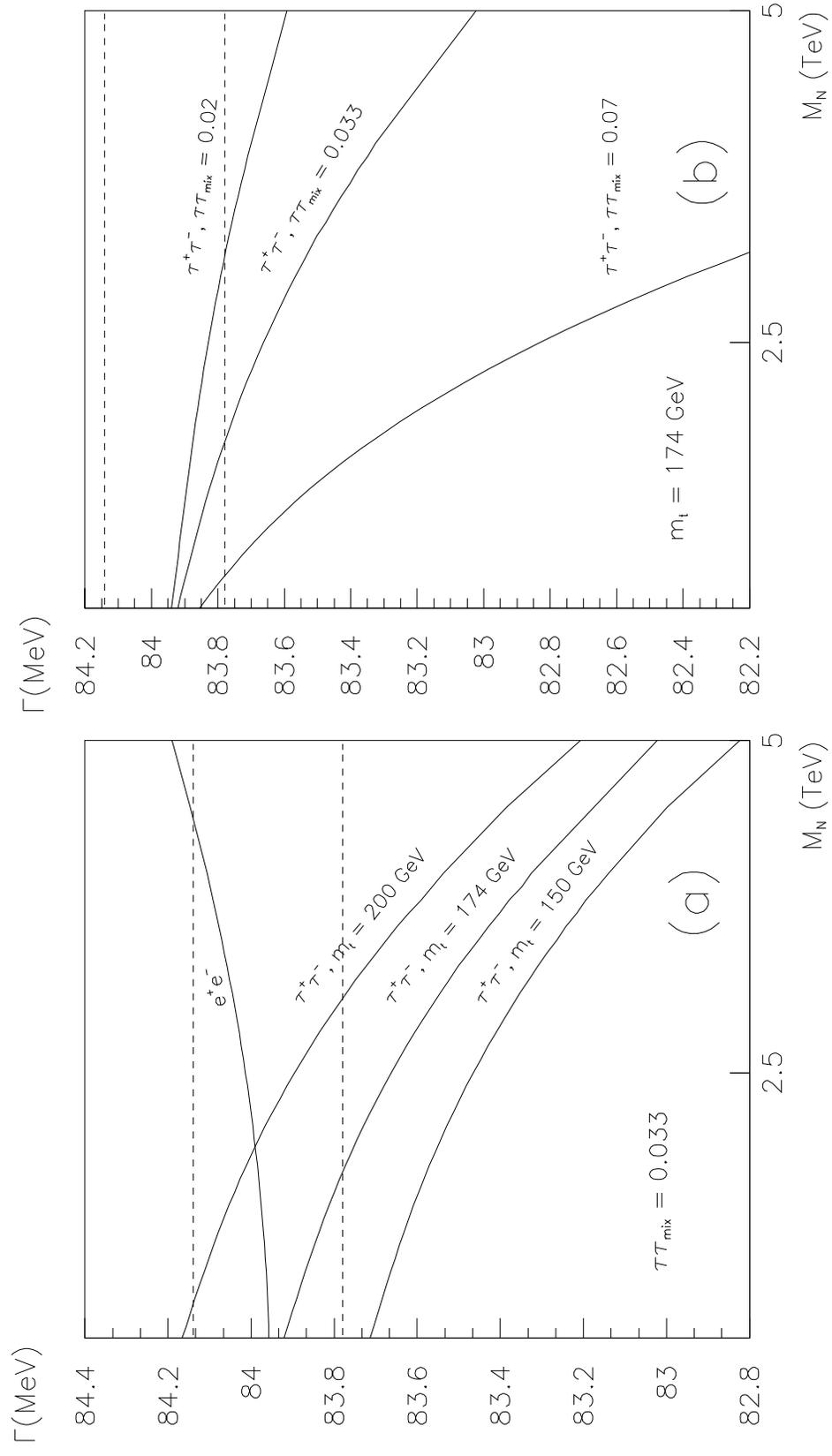
\begin{figure}
\caption{$Z$ leptonic width as a function of $M_{N}$ for (a) fixed mixing
parameter and
different values of $m_{t}$  (b) fixed $m_{t}$ and different values of the
mixing parameter. The dashed lines
represent $1 \sigma$ band about the current experimental value $\Gamma_{l} =
83.96 \pm 0.18 $ MeV.}
\label{fig4}
\end{figure}

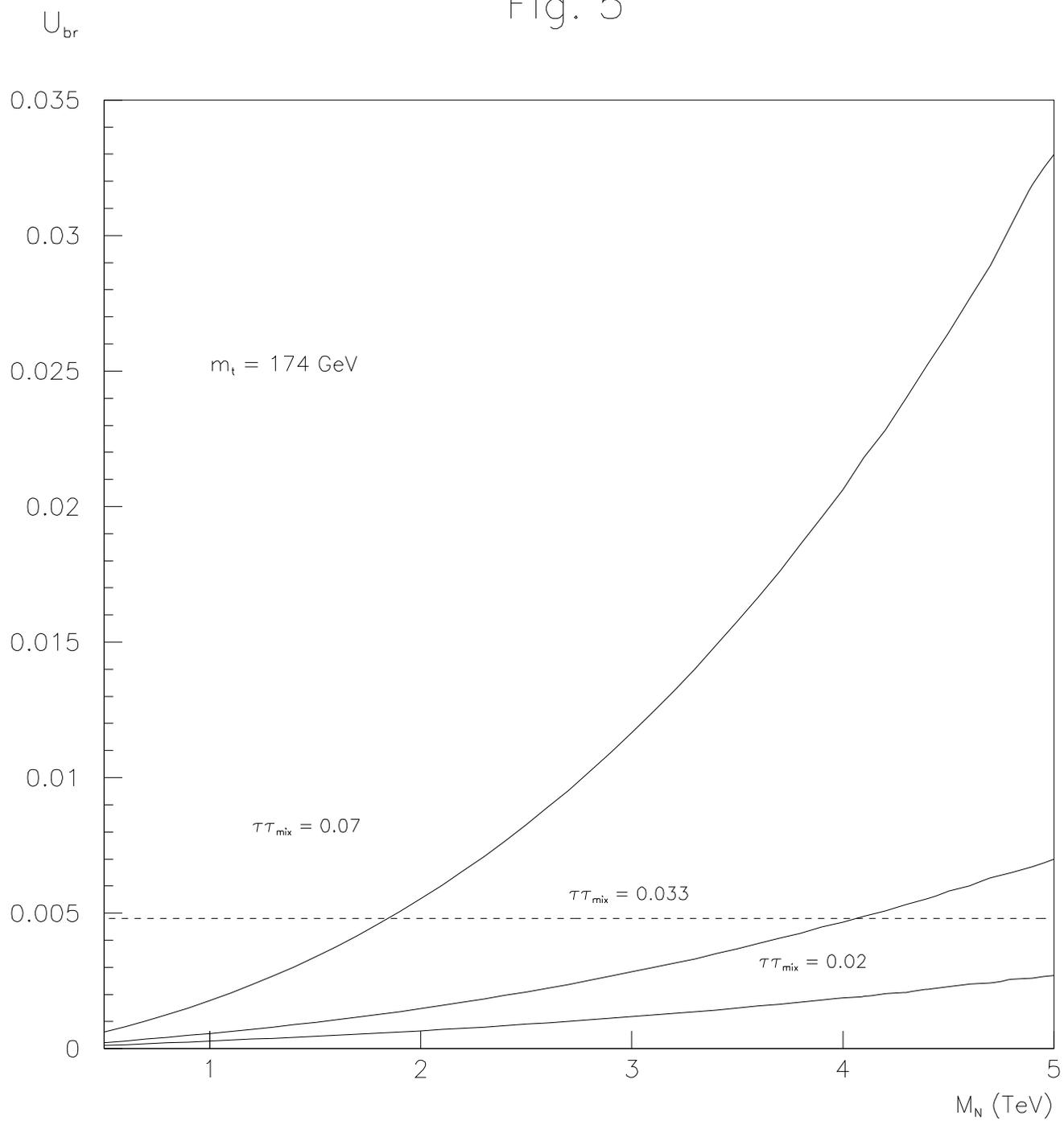
\begin{figure}
\caption{Universality breaking parameter $U_{br}$ as a function of $M_{N}$
for fixed $m_{t}$ and different values of the mixing parameter. The dashed
line represents $1 \sigma$ experimental limit $( < 0.005)$.}
\label{fig5}
\end{figure}

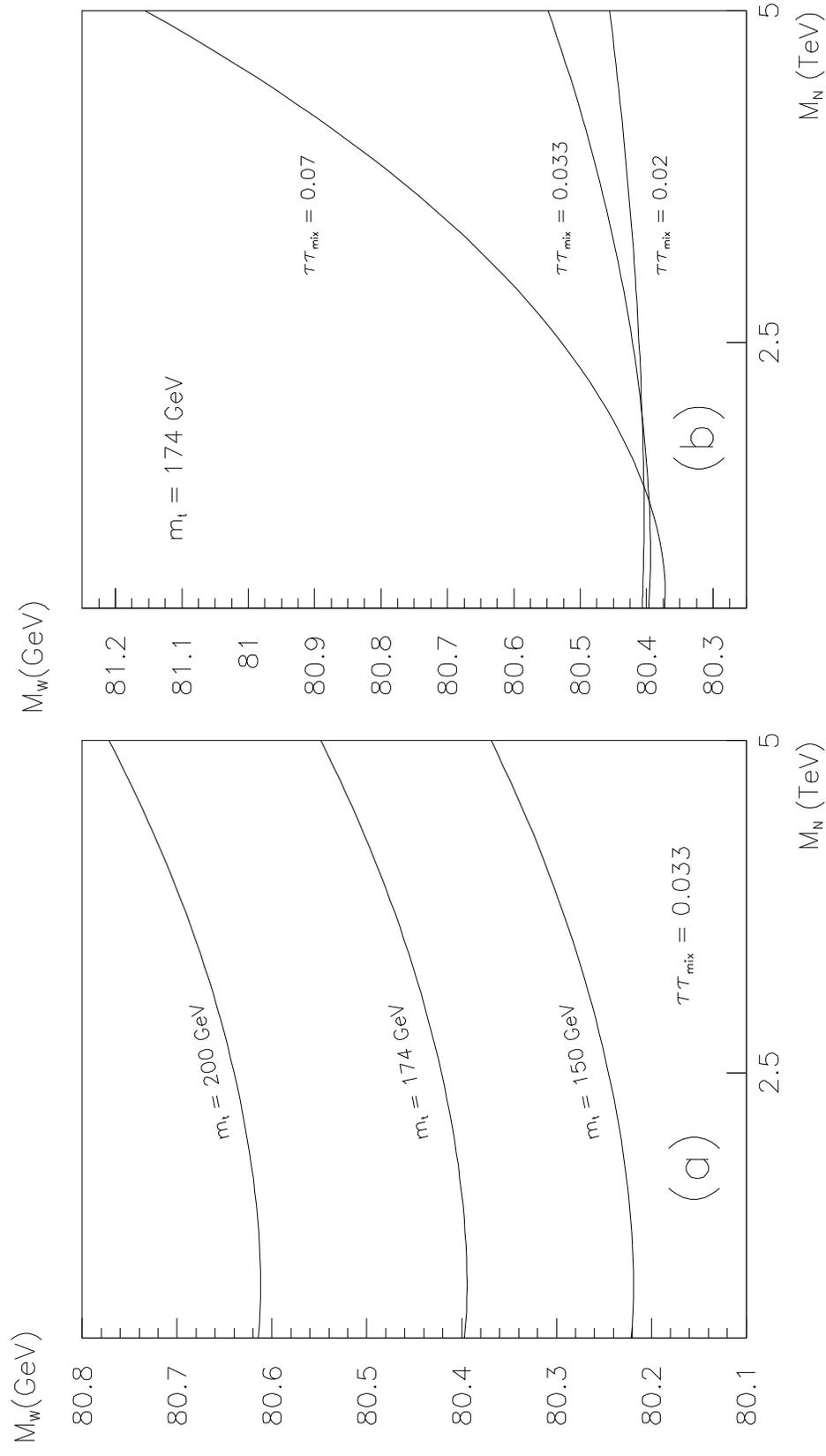
\begin{figure}
\caption{$W$ mass as a function of $M_{N}$
for (a) fixed mixing parameter and different values of $m_{t}$  (b) fixed
$m_{t}$ and different values of the mixing parameter.
The dashed lines represents $1 \sigma$ band about the current
experimental value $M_{W} = 80.23 \pm 0.18$ GeV.}
\label{fig6}
\end{figure}


\end{document}